# Application of Generalized Quaternion in Physics

Liaofu Luo[1] Jun Lv[2]

1, Department of Physics, Inner Mongolia University, Hohhot, 010021, China
Email: lolfcm@imu.edu.cn

2, Department of Physics, Inner Mongolia University of Technology, Hohhot, 010051, China. Email: lujun@imut.edu.cn

Keywords: generalized quaternion, octonion gauge invariance, SU(2) group, $G_2$ group, particle classification

Highlights:

The elementary interaction is deduced from the gauge invariance of the quaternion. The generalizations to SU(3) quaternion and $G_2$ octonion are worked out. The $G_2$ octonion contains seven tri-generator sets of SU(2) symmetry. A model for the elementary particle classification is proposed. It provides a specific approach to unifying the particle interactions.

## Abstract

The applications of quaternion in physics are discussed with an emphasis on elementary particle symmetry and interaction. Three colours of the quark and the quantum chromodynamics (QCD) can be introduced directly from the invariance of basic equations under the quaternion phase transformation (quaternion gauge invariance). The generalized quaternions obey the SU(3) symmetry. QCD is essentially the quantum quaternion dynamics. The further generalization of SU(3) quaternion to $G_2$ octonion is worked out. We demonstrate that the $G_2$ octonion contains seven tri-generator sets of SU(2) symmetry and three of them form SU(3) subgroup. A model of the elementary particle classification and interaction based on octonion gauge theory is proposed. The model unifies the colour and flavour of all particles. It provides a framework for the unified description of four kinds of elementary particles (quarks, leptons, gauge fields and Higgs bosons) and their interactions.

## Text

## 1 Introduction

Quaternion $Z = a + bi + cj + dk$ is the generalization of the complex number $\psi = a + bi$ where $a,b,c$ and $d$ are real numbers and $i$, $j$ and $k$ are the basic quaternions obeying

$$i^2 = j^2 = k^2 = ijk = -1 \tag{1.1}$$

The important feature of a quaternion is that the multiplication does not obey the commutative rule. In 1843 Irish scientist WR Hamilton found the quaternion. [1] Then he tried to find its application in his life but without success. This is an “Irish tragedy”. Once CN Yang said that physics is inseparable from symmetry but the symmetry of quaternions has not been used in physical phenomena. He predicted that quaternions are bound to enter physics and the Irish tragedy would turn into

comedy.[2][3] In the meanwhile, mathematicians believed that "quaternions can supply a shortcut for pure mathematicians who wish to familiarize themselves with certain aspects of theoretical physics"[4]. In recent years it was studied by several authors that various physical covariance groups can all be related to the quaternion group in modern algebra [5]. Quaternions were also widely discussed in applied mathematics [6]，quantitative biology and bioinformatics [7]. However, this article will be limited to the discussion of the application of quaternions and generalized quaternions in particle physics that is related to the restatement of the basic law of particle interactions.

## 2. From complex wave function of the electron to quaternion wave function of quark.

Schrodinger's equation introduced complex numbers into physics. He had attempted to obtain an equation of wave function without complex but he finally gave up the effort. The complex wave function means it contains amplitude and phase to describe the motion of an electron. This is the first reminder of the importance of phase in physics. The invariance of the Schrodinger equation under phase transformation (called gauge transformation) leads to the conservation of electric charge. That is,

$$i\hbar\frac{\partial\psi}{\partial t}=H\psi \tag{2.1}$$

is invariant under gauge transformation $\psi\to\exp(i\theta)\psi$ and this leads to the electric charge conservation from the Noether theorem.

However, there are three phases in a quaternion Z instead of one phase in a complex number $\Psi$. When we introduce quaternion to describe the particle motion there should be new conservation laws. The quaternion Z can be expressed by the Pauli matrix

$$Z = a + i\vec{\mathrm{b}}\cdot\vec{\sigma} = \begin{pmatrix} a+ib_3 & b_2+ib_1 \\ -b_2+ib_1 & a-ib_3 \end{pmatrix}$$

$$\sigma_1 = \begin{pmatrix} 0 & 1 \\ 1 & 0 \end{pmatrix}, \quad \sigma_2 = \begin{pmatrix} 0 & -i \\ i & 0 \end{pmatrix}, \quad \sigma_3 = \begin{pmatrix} 1 & 0 \\ 0 & -1 \end{pmatrix} \qquad (2.2)$$

where $b_{1,2,3}$ and $a$ are real numbers. Easily demonstrate that the matrix expression (2.2) satisfies the quaternion multiplication law. That is, for any two quaternions Z and Z' we have

$$ZZ' = \begin{pmatrix} a+ib_3 & b_2+ib_1 \\ -b_2+ib_1 & a-ib_3 \end{pmatrix}\begin{pmatrix} a'+ib_3' & b_2'+ib_1' \\ -b_2'+ib_1' & a'-ib_3' \end{pmatrix} \qquad (2.3)$$
$$= aa' - \vec{\mathrm{b}}\cdot\vec{\mathrm{b}}' + i(a'\vec{\mathrm{b}} + a\vec{\mathrm{b}}')\cdot\vec{\sigma} - i(\vec{\mathrm{b}}\times\vec{\mathrm{b}}')\cdot\vec{\sigma}$$

Moreover, the inverse of Z can be defined through

$$Z^{-1} = \frac{Z^*}{|Z|^2} = \frac{a - i\vec{\mathrm{b}}\cdot\vec{\sigma}}{a^2+b_1^2+b_2^2+b_3^2} \qquad (2.4)$$

Therefore Z satisfies the group axioms. The quaternions constitute an SU(2)×U(1) group. The phases of quaternion constitute an SU(2) group.

Note that the same quaternion can be expressed by using 4×4 real matrices as follows

$$Z = a\begin{pmatrix} 1&0&0&0\\0&1&0&0\\0&0&1&0\\0&0&0&1 \end{pmatrix} + b_1\begin{pmatrix} 0&-1&0&0\\1&0&0&0\\0&0&0&-1\\0&0&1&0 \end{pmatrix} + b_2\begin{pmatrix} 0&0&-1&0\\0&0&0&1\\1&0&0&0\\0&-1&0&0 \end{pmatrix} + b_3\begin{pmatrix} 0&0&0&-1\\0&0&-1&0\\0&1&0&0\\1&0&0&0 \end{pmatrix} \qquad (2.5)$$

About the application of quaternion in physics SU(2) group can be used to describe the spin or isospin doublet. However, it cannot describe the three-colour symmetry that has been observed in the experiments of particle physics. To explain the three-colour symmetry we should generalize the quaternion group.

We suggest the generalized quaternion ζ defined by

$$\zeta = \alpha + i\sum_{\mu} \beta_\mu \lambda_\mu$$

$$\lambda_1 = \begin{pmatrix} 0 & 1 & 0 \\ 1 & 0 & 0 \\ 0 & 0 & 0 \end{pmatrix}, \quad \lambda_2 = \begin{pmatrix} 0 & -i & 0 \\ i & 0 & 0 \\ 0 & 0 & 0 \end{pmatrix}, \quad \lambda_3 = \begin{pmatrix} 1 & 0 & 0 \\ 0 & -1 & 0 \\ 0 & 0 & 0 \end{pmatrix}$$

$$\lambda_4 = \begin{pmatrix} 0 & 0 & 1 \\ 0 & 0 & 0 \\ 1 & 0 & 0 \end{pmatrix}, \quad \lambda_5 = \begin{pmatrix} 0 & 0 & -i \\ 0 & 0 & 0 \\ i & 0 & 0 \end{pmatrix}, \quad \lambda_6 = \begin{pmatrix} 0 & 0 & 0 \\ 0 & 0 & 1 \\ 0 & 1 & 0 \end{pmatrix}$$

$$\lambda_7 = \begin{pmatrix} 0 & 0 & 0 \\ 0 & 0 & -i \\ 0 & i & 0 \end{pmatrix}, \quad \lambda_8 = \begin{pmatrix} \frac{1}{\sqrt{3}} & 0 & 0 \\ 0 & \frac{1}{\sqrt{3}} & 0 \\ 0 & 0 & \frac{-2}{\sqrt{3}} \end{pmatrix} \qquad (2.6)$$

where ∝ and $\beta_\mu$ are real numbers and $\lambda_1,\ldots\lambda_8$ are generators of the group SU(3). Eq(7) is the direct extension of Eqs(2.1) and (2.2) from the 2×2 Pauli matrix to the 3×3 λ-matrix. Eq(2.6) means the generalized quaternion forming an SU(3)×U(1) group and its phase forming an SU(3) group.

Suppose the dynamical equation of wave function $\Psi_\zeta$ is invariant under quaternion gauge transformation (phase transformation)

$$\psi_\varsigma'(x) = \psi_\varsigma(x) + i(\sum_{\mu}^{8} \beta_\mu \lambda_\mu)\psi_\varsigma(x) \qquad (2.7)$$

we obtain the conservation of colour charge Q= $-i\int \pi \sum_{\mu}^{8} \beta_{\mu}\lambda_{\mu}\psi_{\varsigma} d^3x$

where π is the canonical momentum of $\Psi_{\zeta}$ in the Lagrangian formulation.

## 3. From electromagnetic interaction of electron to chromatic interaction of quark.

As is well known, the electric charge is not only a conservation quantity but also a measure of electromagnetic interaction. Similarly, the colour of a quark is not only a conservation quantity but also a measure of quark's strong interaction. The point can be understood from the gauge transformation. If we generalize the global gauge transformation $\psi \rightarrow \exp(i\theta)\psi$ to local then from the invariance of equation (2.1) under local transformation $\psi \rightarrow \exp(i\theta(x))\psi$ the electromagnetic interaction of electron can immediately be deduced. Similarly, from the invariance of wave equation under local gauge transformation (the generalization of Eq (2.7) )

$$\psi_{\varsigma}'(x) = \psi_{\varsigma}(x) + i(\sum_{\mu}^{8} \beta_{\mu}(x)\lambda_{\mu})\psi_{\varsigma}(x)$$

defined by the generalized quaternion the Yang-Mills interaction of colour quarks is deduced. That is, the existence of SU(3) chromatic interaction, i.e. the existence of gluon interaction between colour quarks, is a logical inference of the symmetry of the generalized quaternion phase transformation.

Historically, quantum chromodynamics (QCD) was firstly published in 1973, nine years after the quark model of hadrons proposed by Gell-Mann and Zweig in 1964. [8][9][10]. We conceive that if the quaternion and its extension were accepted by physics earlier, i.e. the generalized quaternion gauge transformation of quarks was recognized earlier, QCD will be proposed as earlier as in the sixties of the 20th century.

## 4 Further extending quaternions to octonions for studying particle classification and interaction

The quaternions can be extended to octonions in mathematics [11]. We shall discuss the application of octonions in particle physics. Recently Furey tried to use complex octonions to describe the three-generation structure of the standard model of the elementary particles [12][13]. However, Cartan mentioned in 1908 and 1914 that the automorphism group of octonions is a 14-dimensional simple Lie group and that this is the compact real form of $G_2$. [14][15] We will use the compact real form of $G_2$ to study the problem.

Suppose the $G_2$ octonion is defined by 7-dimensional real matrices

$$\zeta=\alpha+i\sum_{\mu=1}^{14}\beta_{\mu}\omega_{\mu} \tag{4.1}$$

$$\omega_1=\begin{bmatrix}0&0&0&0&0&0&0\\0&0&1&0&0&0&0\\0&-1&0&0&0&0&0\\0&0&0&0&-1&0&0\\0&0&0&1&0&0&0\\0&0&0&0&0&0&0\\0&0&0&0&0&0&0\end{bmatrix}\quad
\omega_2=\begin{bmatrix}0&0&-1&0&0&0&0\\0&0&0&0&0&0&0\\1&0&0&0&0&0&0\\0&0&0&0&0&-1&0\\0&0&0&0&0&0&0\\0&0&0&1&0&0&0\\0&0&0&0&0&0&0\end{bmatrix}\quad
\omega_3=\begin{bmatrix}0&1&0&0&0&0&0\\-1&0&0&0&0&0&0\\0&0&0&0&0&0&0\\0&0&0&0&0&0&1\\0&0&0&0&0&0&0\\0&0&0&0&0&0&0\\0&0&0&-1&0&0&0\end{bmatrix}$$

$$\omega_4=\begin{bmatrix}0&0&0&0&-1&0&0\\0&0&0&0&0&1&0\\0&0&0&0&0&0&0\\0&0&0&0&0&0&0\\1&0&0&0&0&0&0\\0&-1&0&0&0&0&0\\0&0&0&0&0&0&0\end{bmatrix}\quad
\omega_5=\begin{bmatrix}0&0&0&1&0&0&0\\0&0&0&0&0&0&-1\\0&0&0&0&0&0&0\\-1&0&0&0&0&0&0\\0&0&0&0&0&0&0\\0&0&0&0&0&0&0\\0&1&0&0&0&0&0\end{bmatrix}\quad
\omega_6=\begin{bmatrix}0&0&0&0&0&0&1\\0&0&0&1&0&0&0\\0&0&0&0&0&0&0\\0&-1&0&0&0&0&0\\0&0&0&0&0&0&0\\0&0&0&0&0&0&0\\-1&0&0&0&0&0&0\end{bmatrix}$$

$$\omega_7=\begin{bmatrix}0&0&0&0&0&-1&0\\0&0&0&0&-1&0&0\\0&0&0&0&0&0&0\\0&0&0&0&0&0&0\\0&1&0&0&0&0&0\\1&0&0&0&0&0&0\\0&0&0&0&0&0&0\end{bmatrix}\quad
\omega_8=\begin{bmatrix}0&0&0&0&0&0&0\\0&0&0&0&0&0&0\\0&0&0&0&0&0&0\\0&0&0&0&1&0&0\\0&0&0&-1&0&0&0\\0&0&0&0&0&0&-1\\0&0&0&0&0&1&0\end{bmatrix}\quad
\omega_9=\begin{bmatrix}0&0&0&0&0&0&0\\0&0&0&0&0&0&0\\0&0&0&0&0&0&0\\0&0&0&0&0&1&0\\0&0&0&0&0&0&1\\0&0&0&-1&0&0&0\\0&0&0&0&-1&0&0\end{bmatrix}$$

$$\omega_{10}=\begin{bmatrix}0&0&0&0&0&0&0\\0&0&0&0&0&0&0\\0&0&0&0&0&0&0\\0&0&0&0&0&0&-1\\0&0&0&0&0&1&0\\0&0&0&0&-1&0&0\\0&0&0&1&0&0&0\end{bmatrix}\quad
\omega_{11}=\begin{bmatrix}0&0&0&0&0&0&0\\0&0&0&0&0&-1&0\\0&0&0&0&0&0&-1\\0&0&0&0&0&0&0\\0&0&0&0&0&0&0\\0&1&0&0&0&0&0\\0&0&1&0&0&0&0\end{bmatrix}\quad
\omega_{12}=\begin{bmatrix}0&0&0&0&0&0&0\\0&0&0&0&0&0&-1\\0&0&0&0&0&1&0\\0&0&0&0&0&0&0\\0&0&0&0&0&0&0\\0&0&-1&0&0&0&0\\0&1&0&0&0&0&0\end{bmatrix}$$

$$\omega_{13}=\begin{bmatrix}0&0&0&0&0&0&-1\\0&0&0&0&0&0&0\\0&0&0&0&1&0&0\\0&0&0&0&0&0&0\\0&0&-1&0&0&0&0\\0&0&0&0&0&0&0\\1&0&0&0&0&0&0\end{bmatrix}\quad
\omega_{14}=\begin{bmatrix}0&0&0&0&0&0&0\\0&0&0&0&1&0&0\\0&0&0&-1&0&0&0\\0&0&1&0&0&0&0\\0&-1&0&0&0&0&0\\0&0&0&0&0&0&0\\0&0&0&0&0&0&0\end{bmatrix}\qquad (4.2)$$

where $\omega_\mu$' s are the generators of the compact real form of $G_2$.

Introduce adjoint generators $\omega'_\mu$ as follows

$$\begin{aligned}&\omega'_8=\omega_8+\omega_1 \qquad \omega'_9=\omega_9+\omega_2 \qquad \omega'_{10}=\omega_{10}+\omega_3\\ &\omega'_{11}=\omega_{11}+\omega_4 \qquad \omega'_{12}=\omega_{12}-\omega_5 \qquad \omega'_{13}=\omega_{13}+\omega_6\\ &\omega'_{14}=\omega_{14}+\omega_7\end{aligned} \qquad (4.3)$$

The generators $\omega_\mu$ ($\mu$=1, ...,14) and $\omega'_\mu$($\mu$=8, ...,14) form a complete set satisfying the closure of the algebraic structure of the group $G_2$. That is, the commutator of any two generators equals some member of the set or zero otherwise. For example,

$$[\omega_1,\omega_2]=-\omega'_{10} \quad [\omega_1,\omega_3]=\omega'_9 \quad [\omega_1,\omega_4]=-\omega'_{12} \quad [\omega_1,\omega_5]=-\omega'_{11} \quad [\omega_1,\omega_6]=\omega_{14} \quad [\omega_1,\omega_7]=\omega'_{13}$$
$$[\omega_1,\omega_8]=0 \quad [\omega_1,\omega_9]=\omega_{10} \quad [\omega_1,\omega_{10}]=-\omega_9 \quad [\omega_1,\omega_{11}]=\omega_{12} \quad [\omega_1,\omega_{12}]=-\omega_{11} \quad [\omega_1,\omega_{13}]=\omega_{14}$$
$$[\omega_1,\omega_{14}]=-2\omega'_{13} \quad [\omega_1,\omega'_8]=0 \quad [\omega_1,\omega'_9]=-\omega_3 \quad [\omega_1,\omega'_{10}]=\omega_2 \quad [\omega_1,\omega'_{11}]=\omega_5 \quad [\omega_1,\omega'_{12}]=\omega_4$$
$$[\omega_1,\omega'_{13}]=2\omega_{14} \quad [\omega_1,\omega'_{14}]=-\omega'_{13}$$

（4.4）

etc. The commutator equal zero occurs only in the following cases

$$[\omega_i,\omega_{i+7}]=0 \quad [\omega_i,\omega'_{i+7}]=0 \quad [\omega_{i+7},\omega'_{i+7}]=0 \quad (i=1,\dots,7) \tag{4.5}$$

Eq (4.4) shows the closeness of the algebraic structure requiring the introduction of adjoint generators $\omega'_\mu(\mu=8, \dots,14)$.

Moreover, in the generator-set of $\omega_\mu$ and $\omega'_\mu$ one obtains relations

$$\begin{aligned}
&\omega_4\omega_7=-\omega_7\omega_4=-\omega'_{10}\\
&\omega_1\omega_{14}=-\omega_{14}\omega_1=-\omega'_{13}\\
&\omega_2\omega'_{12}=-\omega'_{12}\,\omega_2=\omega'_{14}\\
&\omega'_8\,\omega_{11}=-\omega_{11}\omega'_8=\omega_{12}\\
&\omega'_9\,\omega'_{11}=-\omega'_{11}\,\omega'_9=-\omega_{13}\\
&\omega_3\omega_5=-\omega_5\omega_3=-\omega_6\\
&\omega_8\omega_9=-\omega_9\omega_8=-\omega_{10}
\end{aligned} \tag{4.6}$$

etc. and their cycles. For example, in the first of Eq (4.6), the cycles are $\omega_7\omega'_{10}=-\omega'_{10}\,\omega_7=-\omega_4, \quad \omega_{10}'\omega_4=-\omega_4\omega'_{10}=-\omega_7$. Then one finds there exist seven subsets of 3-generators as given in Table 1 and each subset forms an SU(2) subgroup. The generator $\omega_3$, $\omega_5$, or $\omega_6$ of the first set in Table 1 occupies 4 rows and 4 columns of a 7×7 matrix. Namely, all non-zero elements of $\omega_3$, $\omega_5$, and $\omega_6$ are located in row/columns 1-2 and 4-7, 1-4 and 2-7, 1-7 and 2-4 respectively (see Eq (4.2)). Rows 1,2,4 and 7 and columns 1,2,4 and 7 form the reduced 4×4 matrix representations of $\omega_3$ $\omega_5$ and $\omega_6$. Comparing three 4×4 matrices of $\omega_3$ $\omega_5$ and $\omega_6$ with Pauli matrices in Eq (2.5) it is easy to demonstrate that $\omega_3$ $\omega_5$ and $\omega_6$ form the SU(2) subgroup. The other six subsets of 3-generators in Tab 1 obeying SU(2) symmetry can be demonstrated in the same way. It is not difficult to prove that the above seven subsets of SU(2)-generators given in Table 1 form the complete sets of SU(2) symmetry in $G_2$. Therefore, after the introduction of adjoint generators $\omega'_\mu$ ($\mu$=8, ...,14) we can give the representations of SU(2) symmetry. These SU(2) subgroups will play an important role in the discussion of the symmetry and classification of elementary particles.

Table 1 Generator-sets in $G_2$ group with SU(2) symmetry

| **Set** | **Generators with SU(2) symmetry** | **Related row/column** | **SU(2) Relations** |
|---|---|---|---|
| $\mathbf{\Omega_1}$ | $\omega_3, \omega_5, \omega_6$ | 1,2,4,7 | $\omega_3\omega_5=-\omega_6$ and its cycles |
| $\mathbf{\Omega_2}$ | $\omega_8, \omega_9, \omega_{10}$ | 4,5,6,7 | $\omega_8\omega_9=-\omega_{10}$ and its cycles |
| $\mathbf{\Omega_3}$ | $\omega_4, \omega_7, \omega'_{10}$ | 1,2,5,6 | $\omega_4\omega_7=-\omega'_{10}$ and its cycles |
| $\mathbf{\Omega_4}$ | $\omega_1, \omega_{14}, \omega'_{13}$ | 2,3,4,5 | $\omega_1\omega_{14}=-\omega'_{13}$ and its cycles |
| $\mathbf{\Omega_5}$ | $\omega'_8, \omega_{11}, \omega_{12}$ | 2,3,6,7 | $\omega'_8\omega_{11}=\omega_{12}$ and its cycles |
| $\mathbf{\Omega_6}$ | $\omega_2, \omega'_{12}, \omega'_{14}$ | 1,3,4,6 | $\omega_2\omega'_{12}=\omega'_{14}$ and its cycles |
| $\mathbf{\Omega_7}$ | $\omega'_9, \omega'_{11}, \omega_{13}$ | 1,3,5,7 | $\omega'_9\omega'_{11}=-\omega_{13}$ and its cycles |

**$G_2$ invariance and $G_2$ symmetry breaking** There are several $G_2$ invariants

$$\Omega_1^2+\Omega_2^2+\Omega_3^2+\Omega_4^2+\Omega_5^2+\Omega_6^2+\Omega_7^2=-4 \quad (4.7)$$

($\Omega_i$ takes anyone from 3 generators given in Table 1) and

$$\begin{aligned}
&(\omega'_{14}+\omega'_{11})^2=(\omega'_{12}+\omega_{13})^2\\
&(\omega'_{12}+\omega'_{11})^2+(\omega'_{14}+\omega_{13})^2=-4 \quad (between \quad \Omega_6,\Omega_7)\\
&(\omega_{11}+\omega_2)^2=(\omega'_{14}+\omega'_8)^2\\
&(\omega_2+\omega'_8)^2+(\omega'_{14}+\omega_{11})^2=-4 \quad (between \quad \Omega_5,\Omega_6)\\
&(\omega_{14}+\omega_{12})^2=(\omega'_{13}+\omega_{11})^2\\
&(\omega'_{13}+\omega_{12})^2+(\omega_{14}+\omega_{11})^2=-4 \quad (between \quad \Omega_4,\Omega_5)\\
&(\omega'_8+\omega_{13})^2=(\omega_{12}+\omega'_9)^2\\
&(\omega'_8+\omega'_9)^2+(\omega_{12}+\omega_{13})^2=-4 \quad (between \quad \Omega_5,\Omega_7)\\
&(\omega_1+\omega_2)^2=(\omega'_{13}+\omega'_{12})^2\\
&(\omega_1+\omega'_{12})^2+(\omega'_{13}+\omega_2)^2=-4 \quad (between \quad \Omega_4,\Omega_6)\\
&(\omega_1+\omega'_9)^2=(\omega_{14}+\omega'_{11})^2\\
&(\omega_1+\omega'_{11})^2+(\omega_{14}+\omega'_9)^2=-4 \quad (between \quad \Omega_4,\Omega_7)
\end{aligned} \quad (4.8)$$

Eq (4.8) gives pair-correlations between two $\Omega_i$. The relations in Eqs (4.7) and (4.8) can be deduced from the $7\times7$ matrix form in Eq (4.2).

Symmetry breaking is an important issue for the application of $G_2$ symmetry. The $G_2$ symmetry-breaking can be introduced by observing the matrix form of $G_2$ generators. A relation of weak $G_2$ invariance is defined through the reduced $6\times6$ matrix by deleting one column/row of elements 0 in the original matrix. We obtained seven sets of weak $G_2$ invariant relations in the 6-dimensional subspace,

$$\begin{aligned}
&\Omega_2^2+\Omega_4^2+\Omega_5^2=-2, \quad \Omega_2^2+\Omega_6^2+\Omega_7^2=-2, \quad \Omega_1^2+\Omega_2^2+\Omega_3^2=-2\\
&\Omega_3^2+\Omega_5^2+\Omega_7^2=-2, \quad \Omega_1^2+\Omega_5^2+\Omega_6^2=-2, \quad \Omega_1^2+\Omega_4^2+\Omega_7^2=-2\\
&\Omega_3^2+\Omega_4^2+\Omega_6^2=-2
\end{aligned} \quad (4.9)$$

($\Omega_i$ takes anyone from 3 generators given in Table 1). The *a*-th relation (*a*=1,...,7) in Eq (4.9) defines three transformations of $\Omega_i$ leaving the *a*-th basic element in $G_2$ invariant. For example, the first relation defines an SU(3) invariance based on three SU(2) transformations $\Omega_2$, $\Omega_4$ and $\Omega_5$ leaving the first basic element in $G_2$ fixed. This SU(3) subgroup is a broken $G_2$ symmetry with symmetry breaking referred to the “direction” of the first basic element. It means when any disturbance destroys the $G_2$ symmetry of the system the weak $G_2$ invariance can still exist in a subspace as a broken $G_2$ symmetry. Eq (4.9) shows that there exist seven independent ways to define the SU(3) subgroup of $G_2$. The above discussions are consistent with previous references [16] where Gunaydin and Gursey indicated that the isomorphisms of the octonion algebra leaving a basic element invariant form a subgroup SU(3) of $G_2$.

Moreover, the relations

$$\Omega_i^2 = -1 \quad (i=1,\ldots,7) \qquad (4.10)$$

($\Omega_i$ takes anyone from 3 generators given in Table 1) are weak $G_2$ invariants in a 4-dimensional subspace defined through the reduced $4\times4$ matrix by deleting three columns/rows of elements 0 in the original matrix. This is the second kind weak $G_2$ invariance. Since any weak $G_2$ invariance corresponds to a broken $G_2$ symmetry, Eq (4.10) shows there exist seven SU(2) subgroups or seven broken $G_2$ symmetries. It can be proved that no other weak $G_2$ invariants as Eqs (4.9) and (4.10) in the lower subspace. The above two kinds of broken $G_2$ symmetry will provide a wealth of diversity in the particle classification.

## 5 Enlightenment in elementary particle classification and interaction

**5.1 The unification of colour and flavour in $G_2$** In 1974 Georgi and Glashow introduced a grand unified theory of elementary particles based on the SU(5) gauge group which accommodates the standard model of particle physics [17]. However, the extra generators of SU(5) predicted the proton decay rate which conflicts with recent experiments.[18][19]. We have to find another way to unify the elementary particle interactions. Furey demonstrated that the $G_{sm}$ symmetry SU(3)×SU(2)×U(1) can be deduced from the division algebra [12]. The above discussions on the complete sets of SU(2) symmetry in $G_2$ and the broken $G_2$ symmetry provide a starting point and specific approaches to establishing a unified theory of elementary particles without extra hypothetical particles and interactions.

We propose a model in that the elementary particles are classified following the $G_2$ symmetry and broken $G_2$ symmetry. By using any weak $G_2$ invariant relation in Eq (4.9) we define an SU(3) subgroup based on three SU(2) transformations and assume it describes the SU(3) colour transformation for quarks. To be definite, we choose the third relation ($a$=3). It means the broken $G_2$ symmetry constructed from ($\Omega_1$,$\Omega_2$,$\Omega_3$) has been assigned leaving the third basic element invariant. Apart from colour symmetry defined by ($\Omega_1$,$\Omega_2$,$\Omega_3$) we assume the remaining four SU(2) subsets $\Omega_4$ to $\Omega_7$ describe the flavour transformations of particles. From Eq (4.10) we know each of these four transformations satisfies weak $G_2$ invariance. Therefore, the model unifies the colours of quark and the flavours of all elementary particles in a simple $G_2$ group by using two kinds of broken $G_2$ symmetry. The unification of colour and flavour is the basis of the grand unification theory. Since $G_2$ has dimension 14 much lower than dimension 24 of SU(5), the $G_2$ approach of unification is more economic and it can avoid introducing redundant particles and interactions. However, from Eq (4.5) only a few commutators of generators in $G_2$ are equal to zero. It is easy to prove that the SU(3) colour subgroup is not commutative with the SU(2) flavour subgroups. How the colour and flavour can be assigned to a particle simultaneously? Consider the early epoch of particle formation. One may assume the disturbance has destroyed the original $G_2$ symmetry and only the broken $G_2$ symmetry is preserved in the sub-space. The assignment of colour and flavour jointly to a particle is possible because it is based on the existence of two kinds of broken $G_2$ gauge invariance. The colour subgroup and the flavour subgroup are defined in different sub-spaces of $G_2$, the former in 6-dimensional space but the latter in different 4-dimensional spaces.

5.2 **Usages of four subgroups of SU(2) symmetry** After choosing ($\Omega_1$,$\Omega_2$,$\Omega_3$) to construct SU(3) we suggest the remaining four SU(2) subgroups, $\Omega_4$ to $\Omega_7$ ,describe four kinds of flavour symmetries, namely, two flavour symmetries for quarks and leptons respectively and two Higgs doublets. It requires two different SU(2) for quarks and leptons because their adjoint U(1) quantum numbers (hypercharge) are different. However, two local SU(2)×U(1) gauge invariance corresponds to one set of electroweak bosons. The physics behind this usage is: that quarks and leptons have different electric charges but the same electroweak interaction. On the other hand, apart from the first Higgs doublet which was introduced in the Standard Model [20] the second Higgs doublet is necessary for the present theory to fully use seven SU(2) generator-sets of the $G_2$ group. Interestingly, the 2HDM model (two-Higgs-doublet model) [21] has been proven consistent with the recent high-precision measurement of W boson mass and muon magnetic moment that deviated from the Standard Model predictions [22][23]. Detailed theoretical analyses and various experimental predictions of 2HDM in light of the CDF $m_w$ measurement have been proposed [24]. The correlations between four SU(2) doublets obtained in the present model (invariants given in Eq 4.8) are expected to be helpful in further studies on the observational effects of Higgs bosons.

Quarks, leptons and Higgs bosons have different electric charges and spins, but they have the same SU(2) transformation character in their internal degrees of freedom. We have succeeded in classifying them into four SU(2) subgroups of $G_2$. Following this line, if the connection of the internal degrees of freedom with the space-time symmetry of particles is clarified in the future study we will be able to identify the quarks, leptons and Higgs bosons completely.

5.3 **Particle interaction deduced from local gauge invariance** All gauge bosons (gluons, W, Z and photons, etc) emerge in the natural world by the principle of local gauge invariance. The approach from the global gauge invariance generalized to the local can be achieved only when the global gauge invariance is rigorous. Under disturbance the G2 symmetry is destroyed but the weak $G_2$ invariance relations (4.9) and (4.10) are preserved as the broken $G_2$ symmetry in the sub-spaces. The particle interaction should be deduced through these local $G_2$ gauge transformations. The basic interactions include: (1) the SU(3) strong interaction between quarks and gluons that is deduced from the invariance under the local gauge transformation ($\Omega_1$,$\Omega_2$,$\Omega_3$), (2) the SU(2)×U(1) weak- electromagnetic interaction between quarks, leptons, Higgs and W, Z and photons that is deduced from the invariance under the local gauge transformation $\Omega_4$ to $\Omega_7$. In above-mentioned theory the gauge interactions are deduced logically in a minimalist way without introducing redundant interactions that do not exist in experiments. The agreement between theory and experiments shows the point on the broken $G_2$ symmetry as a rigorous symmetry is correct.

Note： Goldstone boson appears necessarily in models exhibiting spontaneous breakdown of continuous symmetries. However, the symmetry among seven SU(2)-generator-sets $\{\Omega_i, i=1,\ldots 7\}$ is discrete. There is no Goldstone boson appearing in the assignment of $\{\Omega_i\}$ to the appropriate particles. As for how $G_2$ symmetry breaking happens, the problem is related to the starting of the universe's evolution and should be waited for further study.

5.4 ***Generation problem in the present model*** The above theory is appropriate to any generation of leptons and quarks. One can study them repeatedly from generation to generation in different particle subspaces, for example, electrons and neutrinons $\nu_e$ , u and d quarks in the first generation, etc. In the present theory，different from the three colours of quarks, the mechanism of appearance of three generations of quarks and leptons is dynamic rather than symmetrical. In works of literature, there was an interesting proposal to introduce complex octonions where occurs alternative SU(3) symmetry to solve the three-generation problem [13]. However, the gauge principle requires that each rigorous symmetry corresponds to an interaction. Since there is no evidence of the existence of such an interaction, we are inclined to the view that the origin of three generations is dynamic rather than symmetrical. Moreover, the problem of generation repetition is closely related to the mechanism of generation mixing. A typical example of generation mixing is neutrino oscillation which means the violation of the conservation of individual leptonic numbers and has been considered to be evidence for physics beyond the standard model. However, the neutrino oscillation experiments have shown that neutrinos do have mass and the problem of neutrino mass can be solved in 2HDM by considering the one-loop correction (the neutrino of the i-th generation oscillating to the j-th generation coupled with two-Higgs-doublet) in the parameter calculation [22]. It supports the view of the dynamic origin of generation mixing.

## 6 Conclusions

Through the extension of the symmetry from quaternion to SU(3) quaternion, and to $G_2$ octonion we demonstrated that the octonion contains seven SU(2)-generator-sets. Based on 7 SU(2) subgroups and two kinds of broken $G_2$ gauge invariance, a model for the elementary particle classification and interaction is proposed. The model unifies the colours and flavours of all elementary particles and unifies the SU(3) strong interaction and SU(2)×U(1) weak-electromagnetic interaction in a minimalist way. Moreover, the proposed model supports the version of two Higgs doublet that can explain the new physics beyond the Standard

Model.

Acknowledgement: The authors thank Dr Furey for her work on the usage of division algebra in particle physics and CDF collaboration for their high-precision measurement of the standard model deviation. It was their outstanding work that inspired the research of this project.

## References

1) WR Hamilton. On quaternions or on a new system of imaginaries in algebra. Phylosophical Magazine 25(3) 489-495 (1844).

2) CN Yang. The separation and combination of mathematics and physics in the past hundred years. Global Science No 10 (2008) (in Chinese).

3) DZ Zhang. Talk with CN Yang on the relation between mathematics and physics. Mathematical Intelligencer. v15, No4 (1993).

4) J Lambek. If Hamilton prevailed: quaternions in physics. Mathematical Intelligencer. v17, No4 (1995).

5) PR Girared. The quaternion group and modern physics. European Journal of Physics 5(1) 25-32 (1984).

6）K Karsten, S Helmut. The Bingham distribution of quaternions and its spherical radon transform in texture analysis.

Mathematical Geology. **36** (8): 917–943 (2004).

7) JJ Shu, Y Li. Hypercomplex cross-correlation of DNA sequences. J Biol. Systems 18(4) 711-725 (2010).

8) M Gell-Man. A schematic model of baryons and mesons. Physics Letters 8 (3) 214-215 (1964).

9) G Zweig. An SU(3) model for strong interaction symmetry and its breaking. CERN Report No. 8182/TH. 401 (1964).

10) H Fritzsch, M Gell-Man, H Leutwyler. Advantages of the color octet gluon picture. Physics Letters B47 (4)365 (1973).

11) JH Conway, DA Smith. On quaternions and octonions: their geometry arithmetic and symmetry. AK Piters ISBN 1-56881-134-9. (2003)

12）C Furey. SU(3)×SU(2)×U(1) as a symmetry of division algebraic ladder operators. Eur. Phys. J C78:375 (2018).

13) C Furey. Three generations, two unbroken gauge symmetries and one eight-dimensional algebra. Physics Letters B 785, 84-89 (2018).

14) E Cartan. Nombres complexes. Encyclopedie des Sciences Mathematiques. Paris: Gauthier-Villars. 329-468 (1908).

15) E Cartan. Les groups reels simples finis et continus. Ann. Sci. Ecole Norm. Sup. 31, 255-262 (1914).

16) M Gunaydin and E Gursey. Quark structure and the octonions. J Math Phys. 14(11) (1973).

17) H Georgi, S Glashow. Unity of all elementary particle forces. Phys

Rev Lett. 32 (8) 438-441(1974).

18) The Super-Kaminokande Collaboration, Search for protein decay via p→νK using 260 kiloton-year data of Super-Kaminokande. Phys. Rev D90, 072005 (2014).

19) The Super-Kaminokande Collaboration, Search for protein decay via p→eπ and p→μπ in 0.31 megaton-years exposure of the Super-Kaminokande water Cherenkov detector (2016). arXiv:1610.03597 [hep-ex].

20) M E Peskin and DV Schroeder. An Introduction to Quantum Field Theory. Addison-Wesley, Reading, 1995. 2nd edition, pbk. Westview P ress. 2015.

21) GC Branco, PM Ferreira, L Lavoura, MN Rebelo, M Sher, and JP Silva. Theoryand phenomenology of two-Higgs-doublet models. Physics Reports. **516** (1): 1–102. ( 2012)

22）TA Chowdhury，J Heeck, S Saad and A Thapa, W boson mass shift and muon magnetic moment in the Zee model. arXiv: 2204.08390.

23) CDF COLLABORATION High-precision measurement of the *W* boson mass with the CDF II detector. Science 376 (6589) 170-176. (Apr 2022). Doi: 10.1126/science.abk1781.

24）S Lee，K Cheung, J Kim, CT Lu and J Song. Status of the two-Higgs-doublet model in light of the CDF $m_W$ measurement, arXiv : 2204.10338.